\documentclass[twocolumn]{emulateapj}

\usepackage{graphicx}
\usepackage{subfigure}
 \pdfoutput=1
\usepackage{latexsym,amssymb}
\usepackage{amsmath}
\usepackage{natbib}

\usepackage[colorlinks=true,linkcolor=blue,citecolor=blue]{hyperref}

\shorttitle{sGRB Jets from Neutron Star Mergers}

\shortauthors{Murguia-Berthier et al.}

\begin{document}

\title{The Properties of Short gamma-ray burst Jets Triggered by neutron star mergers}

\begin{abstract}
The most popular model for short gamma-ray bursts (sGRBs) involves the coalescence of binary neutron stars. Because the progenitor is actually hidden from view, we must consider under which circumstances such merging systems are capable of producing a successful sGRB. Soon after coalescence,  winds are launched from the merger remnant. In this paper, we use realistic wind profiles derived from global merger simulations in order to investigate the interaction of sGRB jets with these winds using numerical simulations. We analyze the conditions for which these axisymmetric winds permit relativistic jets to breakout and produce a sGRB. We find that jets with luminosities comparable to those observed in sGRBs are only successful when their half-opening angles are below $\approx 20^\circ$. This jet collimation mechanism leads to a simple physical interpretation of the luminosities and opening angles inferred  for sGRBs. If wide, low luminosity jets are observed, they might be indicative of a different progenitor avenue such as the merger of a neutron star with a black hole. We also use the observed durations of sGRB to place constraints on the lifetime of the wind phase, which is determined by the time it takes the jet to breakout. In all cases we find that the derived limits argue against completely stable remnants for binary neutron star mergers that produce sGRBs.
\end{abstract}

\author{Ariadna Murguia-Berthier\altaffilmark{1}, Enrico Ramirez-Ruiz\altaffilmark{1}, Gabriela Montes\altaffilmark{1}, Fabio De Colle\altaffilmark{2}, Luciano Rezzolla\altaffilmark{3,4}, Stephan Rosswog\altaffilmark{5}, Kentaro Takami\altaffilmark{3,6}, Albino Perego\altaffilmark{7} and William H. Lee\altaffilmark{8}}
\altaffiltext{1}{Department of Astronomy and
 Astrophysics, University of California, Santa Cruz, CA
 95064}
 \altaffiltext{2}{Instituto de Ciencias Nucleares, Universidad Nacional Aut{\'o}noma de M{\'e}xico, A. P. 70-543 04510 D. F. Mexico}
\altaffiltext{3}{Institute for Theoretical Physics, Goethe University, Max-von-Laue-Str. 1, 60438 Frankfurt am Main, Germany}
\altaffiltext{4}{Frankfurt Institute for Advanced Studies, Ruth-Moufang-Str. 1, 60438 Frankfurt am Main, Germany}
\altaffiltext{5}{Astronomy and Oskar Klein Centre, Stockholm University, AlbaNova, SE-106 91 Stockholm, Sweden}
\altaffiltext{6}{Kobe City College of Technology, 651-2194 Kobe, Japan}
\altaffiltext{7}{Institut f\"{u}r Kernphysik, Technische Universit\"{a}t Darmstadt, D-64289
Darmstadt, Germany}
\altaffiltext{8}{Instituto de Astronom\'ia, Universidad Nacional Aut{\'o}noma de M{\'e}xico, A. P. 70-264 04510 D. F. Mexico}

\section{Introduction}\label{int}
Neutron star binary mergers (NSBMs) are sources of gravitational waves, as well as the most discussed model for short $\gamma$-ray bursts \citep[sGRBs;][]{eichler89, paczynski91, narayan92}. As the binary coalesces, the resulting object depends on the total mass of the merger. If that mass is less than the maximum mass allowed by rigid rotation \citep[constrained to be at least $\approx 2 M_\odot$ by observations of PSR J0348+0432 and PSR J1614-2230;][]{demorest10, antoniadis2013}, it can result in a supramassive neutron star. Furthermore, if it is greater than that threshold mass, the merger will become a hot, differentially rotating, hyper-massive neutron star (HMNS) surrounded by an accretion disk \citep[e.g.][]{baiotti2008}. The fate of the HMNS is uncertain, as it can live stably for a long time, or undergo collapse to a black hole (BH) \citep{shibata06,baiotti2008, ravi2014}. 

In the latter case, delayed collapse to a BH and significant mass loss can occur after sufficient angular momentum transport from the inner to the outer regions of the remnant. Several processes can act to transport angular momentum and drive collapse, including gravitational waves, neutrino-driven winds and magnetic fields \citep{lee07}. As the mass is accreted to the central object, jets can be launched from the compact object. Details about that process remain unsure, though, as central engine models often invoke either prompt collapse to a BH \citep[e.g.][]{ari2014} or the formation of a rapidly spinning, highly magnetized HMNS \citep[e.g.][]{zhang2001, metzger2008magnetar, rezzolla2015}. 

In order to discriminate between progenitor scenarios, we need a better understanding of the way in which the emerging relativistic jet propagates through the surrounding medium \citep{mochkovitch93, rosswog03, rosswog2003b,aloy05, aloy2006, rezzolla11,palenzuela2013, ruiz2016}. There are many collimation mechanisms (and potential death traps) for relativistic jets \citep{lee07} in a NSBM context, including dynamically ejected material from the tidal tails as well as various types of baryon-loaded winds produced during the HMNS phase \citep{siegel2014, perego2014,hotokezaka2013,nagakura2014, duffell2015, sekiguchi2016}. 
In this {\it Letter} we explore the consequences that neutrino-driven and magnetically-driven winds produced during the HMNS phase have on the jet collimation and on determining under which conditions a jet can break out successfully. We use latitudinal density profiles taken from simulations of NSBMs in order to calculate a more realistic circumburst environment.

The {\it Letter} is structured as follows. Section~\ref{ana} gives a description of the properties of winds derived from global NSBMs simulations and investigates how they can potentially alter the properties of the emerging jet. Section~\ref{num} describes the numerical methods used and presents a description of the jet's interaction, thought to be produced after the collapse to a BH, with the previously ejected wind material. Finally, in Section~\ref{dis} we discuss the results obtained and compare them with observations of sGRBs.

\section{Jet advancement and collimation}\label{ana}
The properties of the external environment could hamper the jet's advancement. The jet should be able to break free from the baryon-loaded wind if the velocity of the jet's head is larger than that of the wind material. By balancing momentum fluxes at the jet's working surface, we can obtain \citep[e.g.,][]{begelman1989,bromberg11}:
\begin{equation}
\beta_{\rm h}=\frac{\beta_{\rm j}}{1+\tilde{L}^{-1/2}}
\label{eq:lstar}
\end{equation}
where $\beta_{\rm h}$ and $\beta_{\rm j}$ represent the velocities of the jet's head and the shocked material, respectively. Here
\begin{equation}
\tilde{L}=\frac{\rho_{\rm j}}{\rho_{\rm w}}h_{\rm j}\Gamma_{\rm j}^2,
\label{eq:deflstar}
\end{equation}
where $h_{\rm j}$ is the specific jet's enthalpy, and $\Gamma_{\rm j} $ the initial jet's Lorentz factor, $\rho_{\rm j}$ and $\rho_{\rm w}$ the jet's  and wind's densityespectively. $\tilde{L}$ is the jet's critical parameter that determines the
evolution \citep{bromberg11}, while collimation can be attained if: 
\begin{equation}
\tilde{L}<\theta_{\rm j}^{-4/3},
\label{eq:collimation}
\end{equation}
where $\theta_{\rm j}$ is the jet's half-opening angle. For a jet with axial symmetry, its structure
can be described by the angular distribution of its luminosity content per unit solid angle, $L_\Omega (\theta)$. For this discussion, we shall assume the jet is uniform, where $L_\Omega$ and $\Gamma_{\rm j}$
are constant within $\theta_{\rm j}$ and sharply decreasing at larger polar angles. This means $L_\Omega$ here is the isotropic equivalent luminosity as, for example, derived from the $\gamma$-ray flux. 
The complex nature of the HMNS close to critical rotation leaves open the possibility of the emerging uniform jet interacting with a non-spherical mass distribution: $\rho_{\rm w}(\theta)$. As we show in Section~\ref{winds}, the jet is likely to encounter a slower and denser wind confined to the equatorial plane. To compute the exact latitudinal dependence of the wind properties of HMNS  we require global simulations that reproduce, as realistically as possible, the conditions expected in NSBMs. It is to this problem that we now turn our attention.

\subsection{Constraints derived from neutrino-driven and magnetically-driven winds}\label{winds}
%The dynamical timescale associated with the HMNS is given by
%\begin{equation}
%t_{\rm dyn} \approx 0.14 \left({R_{\rm ns} \over 25\ {\rm km}}\right)^{3/2} \left({M \over 3\ M_\odot}\right)^{-1/2}\;{\rm ms}.
%\label{eq:dyn}
%\end{equation}
Considering the spectral frequencies of the gravitational wave signal of a binary neutron star merger \citep{rezzolla2016}, we can determine a dynamical timescale associated with the rotation of the HMNS to be $t_{\rm dyn} \approx 1\rm{ms}$.
Given the various stabilizing mechanisms, the HMNS is likely to survive for a timescale longer than $t_{\rm dyn}$. Thermal support in the HMNS is governed by neutrino diffusion and the associated cooling time can be written as \citep{perego2014}
\begin{equation}
t_{\rm \nu}\approx1.88 \left({R_{\rm ns} \over 25\ {\rm km}}\right)^{2} \left({\rho_{\rm ns} \over 10^{14}\,{\rm g\,cm^{-3}}}\right) \left({k_{\rm B} T_{\rm ns} \over 15\ {\rm MeV}}\right)^{2}{\rm s}.
\label{eq:cooling}
\end{equation}

Stabilization of HMNS due to differential rotation is expected to last for many $t_{\rm dyn}$ and is presumed to be halted by some dissipative mechanism, like viscosity, gravitational radiation magnetic amplification and/or magnetic braking \citep{price2006, baiotti2008,giacomazzo2011,kiuchi2012,siegel2013, kiuchi2015}. The characteristic timescale for magnetic braking of differential rotation by toroidal Alfv\'en waves is, for example, estimated to be of the order of $R/v_{\rm A}$ \citep{shapiro2000}:
\begin{equation}
t_{\rm A} \approx 0.17 \left({R_{\rm ns} \over 25\ {\rm km}}\right)^{-1/2} \left({B \over 10^{15}\ {\rm G}}\right)^{-1} \left({M_{\rm ns} \over 3\ M_\odot }\right)^{1/2}{\rm s},
\label{eq:alfven}
\end{equation}
where differential rotation has been assumed here to convert an appreciable fraction of the kinetic energy in differential motion in the (initially weakly magnetized) HMNS into magnetic field energy \citep[e.g.][]{siegel2014}. These angular momentum transport processes push the HMNS to uniform rotation, which could lead to its catastrophic collapse to a BH on a timescale $\ll t_{\nu}$ if the excess mass can not be supported \citep{fryer2015, cole2015}.
 
During the HMNS phase, these various dissipation and transport mechanisms give rise to significant mass loss. Because of the density and velocity structure in the HMNS, the accompanying mass loss is expected to be anisotropic \citep[e.g.][]{rosswog03}. The result is a remnant wind that originated in the presence of the HMNS. Though magnetic fields and neutrino interactions likely play a crucial role for the understanding the mass-shedding properties of HMNS, the numerical tools needed to study this problem have not been available until recently. Here we make use of the results of two global simulations of NSBMs aimed at, as realistically as possible, quantifying the properties of neutrino-driven \citep{perego2014} and magnetically-driven \citep{siegel2014} winds from HMNS, respectively.

 \begin{figure}[t!]
\includegraphics[height=0.70\linewidth,clip=true]{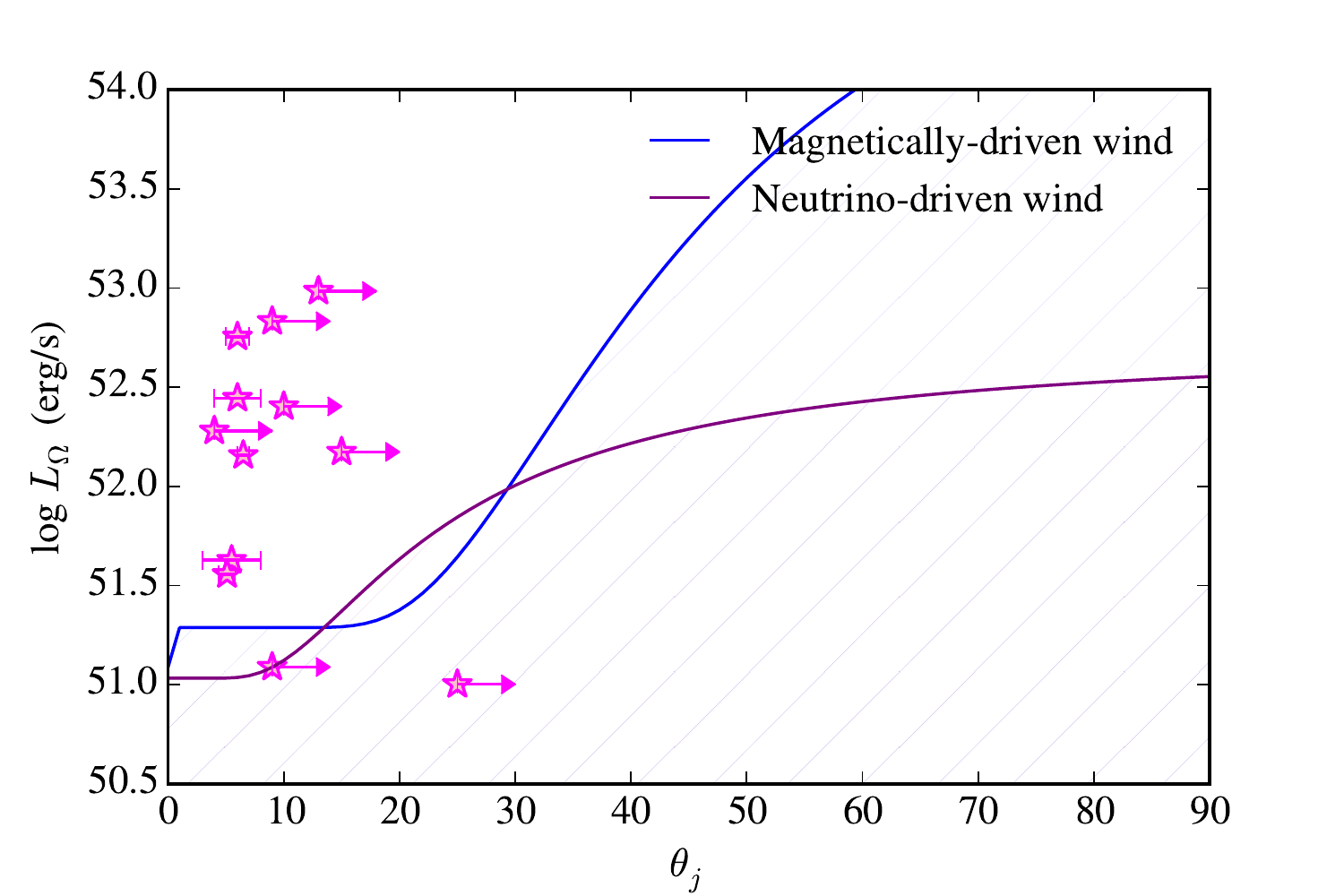}
\caption{$L_{\rm \Omega}$ of a jet, uniform within $\theta_j$, needed to have a successful sGRB. Below the line (dashed region), the jet will choke (i.e. $\beta_{\rm h}<\beta_{\rm w}$). The angular structure of the neutrino-driven and magnetically-driven winds are derived here from \citet{perego2014} and \citet{siegel2014}, respectively. The jet and winds are characterized by $\Gamma_{\rm j}=10$ and $\beta_{\rm w}=0.3$. The symbols represent the lower limit of $L_{\gamma, \rm{iso}}$ from \citet{fong2015, fong2016,troja2016}.}
\label{fig:lvstheta}
\end{figure}	

The angular dependence of the winds are derived by averaging the latitudinal profiles at the end of the simulation in \citet{perego2014}, and relative to the $\tt rand$ configuration of \citet{siegel2014}, and assuming that the wind structure does not evolve with time. It is used here to calculate the minimum $L_{\rm\Omega}$ needed for a sGRB jet to break free from the surrounding wind (i.e. $\beta_{\rm h} >\beta_{\rm w}$). The resulting constraints are plotted in Figure~\ref{fig:lvstheta}. For comparison, the distributions of isotropic equivalent $\gamma$-ray luminosities ($L_{\gamma, \rm{iso}}=E_{\gamma, \rm{iso}}/t_{90}$) and half-opening angles from \citet{fong2015, fong2016,troja2016} are also plotted. It is notable that the constraints derived from both sets of global simulations are rather similar near the burst and they are in remarkable agreement with observations, suggesting that the properties of sGRB jets are likely shaped by the character of the wind emanating from the HMNS.

\begin{figure}[t]
\includegraphics[scale=0.60]{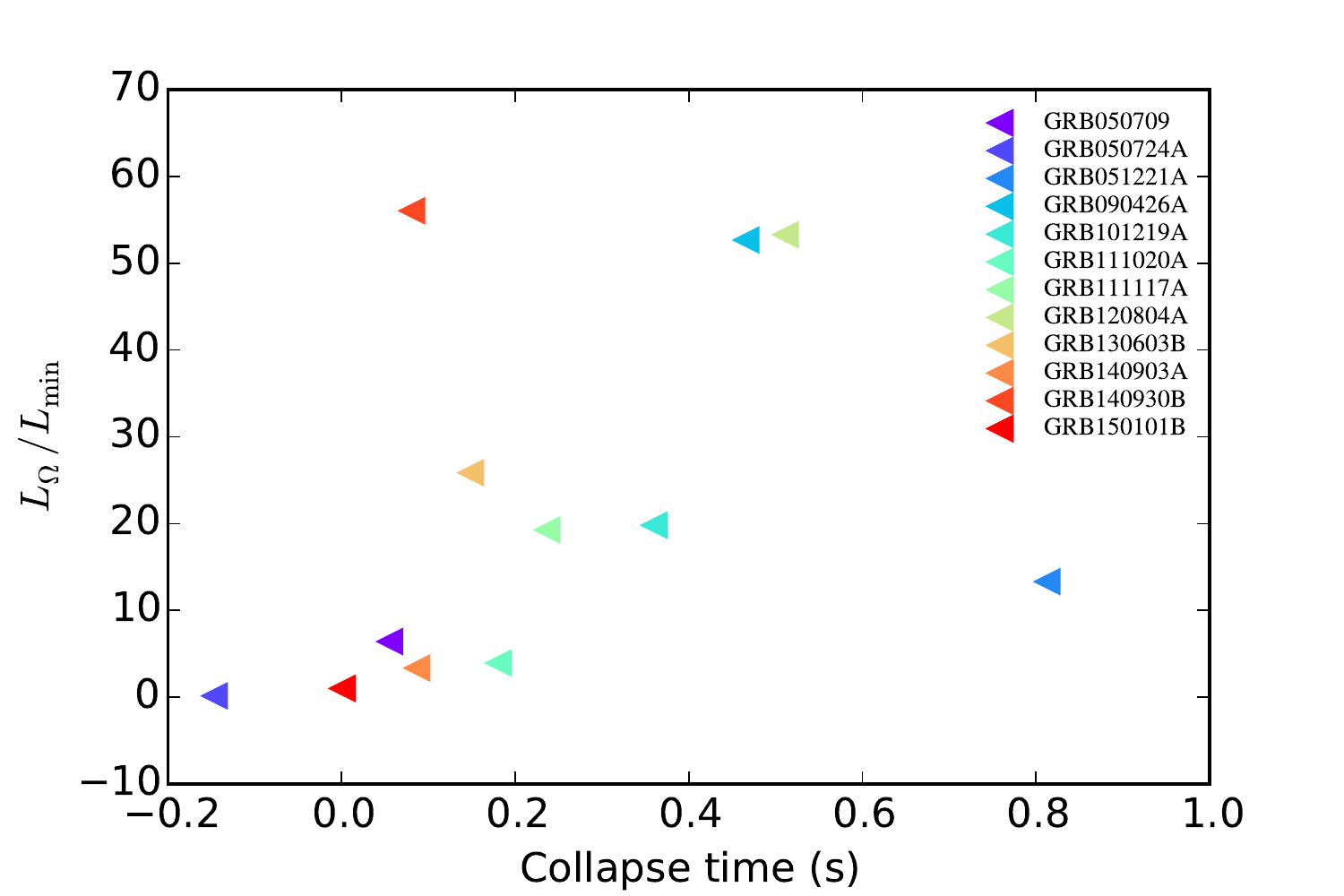}
\caption{Predicted upper limit on the lifetime of the HMNS, $t_{\rm w}$. The symbols represent the limits derived using data from \citet{fong2015, fong2016,troja2016}. Here $L_{\rm min}$ represents the minimum luminosity a jet with a given initial half-opening angle needs to break free from the wind, i.e. below the line in Figure~\ref{fig:lvstheta}. The outlier that has a negative time is GRB050724A, whose luminosity is below $L_{\rm min}$. In Section~\ref{dis} we argue that it could have been produced by the merger of a NS with a BH.}
\label{fig:timecollapse}
\end{figure}

In addition to constraining the luminosity and opening angle needed to produce a successful sGRB, we can derive a limit on the duration of the wind injection phase $t_{\rm w}$, which is determined by the time it takes the HMNS to collapse to a BH. The velocity of the jet's head is subrelativistic while traversing the wind. This means that if the central engine stops the energy input in the jet's head before it breaks free, the head will stay subrelativistic and there will be no emission. The sGRB is successful if:
\begin{equation}
t_{\rm w} \leq t_{\rm sgrb} \frac{\beta_{\rm h}-\beta_{\rm w}}{\beta_{\rm w}},
\label{eq:timecollapse}
\end{equation}
where $ t_{\rm sgrb}$ is the duration of the event.
As long as the condition in equation~(\ref{eq:timecollapse}) is satisfied, the jet will be able to produce a sGRB lasting $\approx t_{\rm sgrb}$. Figure~\ref{fig:timecollapse} shows the limits on $t_{\rm w}$ for the sample compiled by \citet{fong2015, fong2016, troja2016} derived by assuming that such jets need to successfully break free from the HMNS wind (Figure~\ref{fig:lvstheta}). In all cases, the required limits are $\lesssim t_\nu$ (equation~(\ref{eq:cooling})), which argues against the complete stabilization of the HMNS and suggests that the collapse to a BH occurs on a timescale $\approx t_{\rm A} \gg t_{\rm dyn}$ \citep[see e.g.][]{ari2014}. 

\section{Numerical Study}\label{num}
We performed 2D numerical simulations to quantify how the interaction of the wind, ejected during the HMNS phase, modifies the propagation of a relativistic jet, assumed to be produced after the collapse to a BH. The simulations were performed using the \emph{Mezcal} special relativistic hydrodynamic code. It uses adaptive mesh refinement in order to resolve the propagation of relativistic, collimated flows. A description can be found in \citet{decolle2012} along with numerical tests.

\begin{figure}
\centering
\includegraphics[height=1.5\linewidth,clip=true]{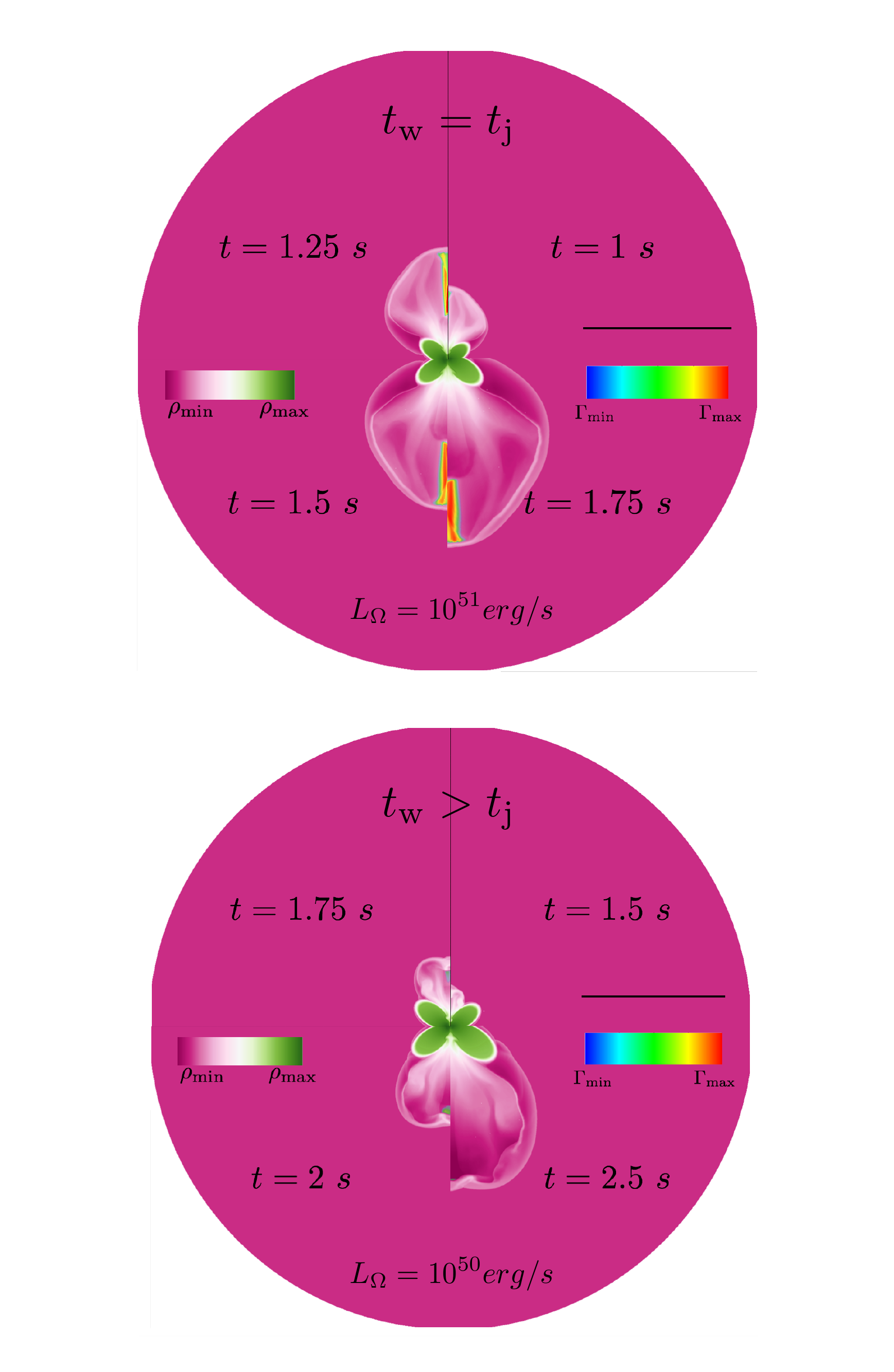}
\caption{The temporal evolution of a jet propagating through a realistic wind. {\it Top} panel: The jet has $L_\Omega=10^{51}{\rm erg\ s^{-1}}$, $\Gamma_{\rm j}=10$, $t_{\rm j}=0.5{\rm{s}}$, and $\theta_{\rm j}=10^\circ$. For the wind: $t_{\rm w}=0.5\rm{s}$. {\it Bottom} panel:  The jet has $L_\Omega=10^{50}{\rm erg\ s^{-1}}$, $\Gamma_{\rm j}=10$, $t_{\rm j}=0.5\rm{s}$, and $\theta_{\rm j}=10^\circ$, For the wind: $t_{\rm w}=1\rm{s}$. Shown are logarithmic density and Lorentz factor contours at different times, where $[\rho_{\rm min}, \rho_{\rm max}]=[7.6\times10^{-7},1386.3]{\rm g\ cm^{-3}}$ and $[\Gamma_{\rm min}, \Gamma_{\rm max}]=[8.0,10.0]$. A $3\times10^{10}\rm{cm}$ scale bar is shown. Calculations were done in 2D spherical coordinates using an adaptive grid of size $l_{\rm r}=6\times10^{10}\rm{cm}$, $l_{\theta}=\frac{\pi}{2}$, with $100\times40$ initial cells, and 5 levels of refinement (maximum resolution of $3.75\times10^{7}\rm{cm}$).}
\label{fig:numerical1}
\end{figure}

The simulation setup follows that implemented by \citet{ari2014}. It begins with the injection of a slow ($\beta_{\rm w}=0.3$), dense wind lasting for  $t_{\rm w}$. After $t_{\rm w}$, the density of the wind is decreased as $t^{-5/3}$ \citep[e.g.][]{lee07} and a jet is introduced. The jet is uniform and characterized by its luminosity ($L_\Omega$), half-opening angle ($\theta_{\rm j}$), Lorentz factor ($\Gamma_{\rm j}=10$) and duration ($t_{\rm j}$). We relax the assumption in \citet{ari2014} of a uniform wind and explore the interaction with a non-spherical distribution: $\rho_{\rm w}(\theta)$, calculated using the same dependance as Figure~\ref{fig:lvstheta}. As shown in the figure, a denser wind will likely impede the jet's advancement.

Initially, the jet is unable to move the wind material at a speed comparable to its own and thus
is decelerated to a Lorentz factor $\Gamma_{\rm h}\ll\Gamma_{\rm j}$. Most of the excess energy is accumulated within a cocoon that engulfs the jet \citep{ramirez-ruiz2002}. If the jet produced by the accretion onto the newly formed BH maintains its luminosity for longer than it takes the jet's head to reach the edge of the wind, a successful sGRB will be produced (as in the case depicted in the {\it top} panel of Figure~\ref{fig:numerical1}). If the jet activity terminates beforehand, the 
jet will be choked ({\it bottom} panel of Figure~\ref{fig:numerical1}).

\begin{figure}
\centering
\includegraphics[height=0.78\linewidth,clip=true]{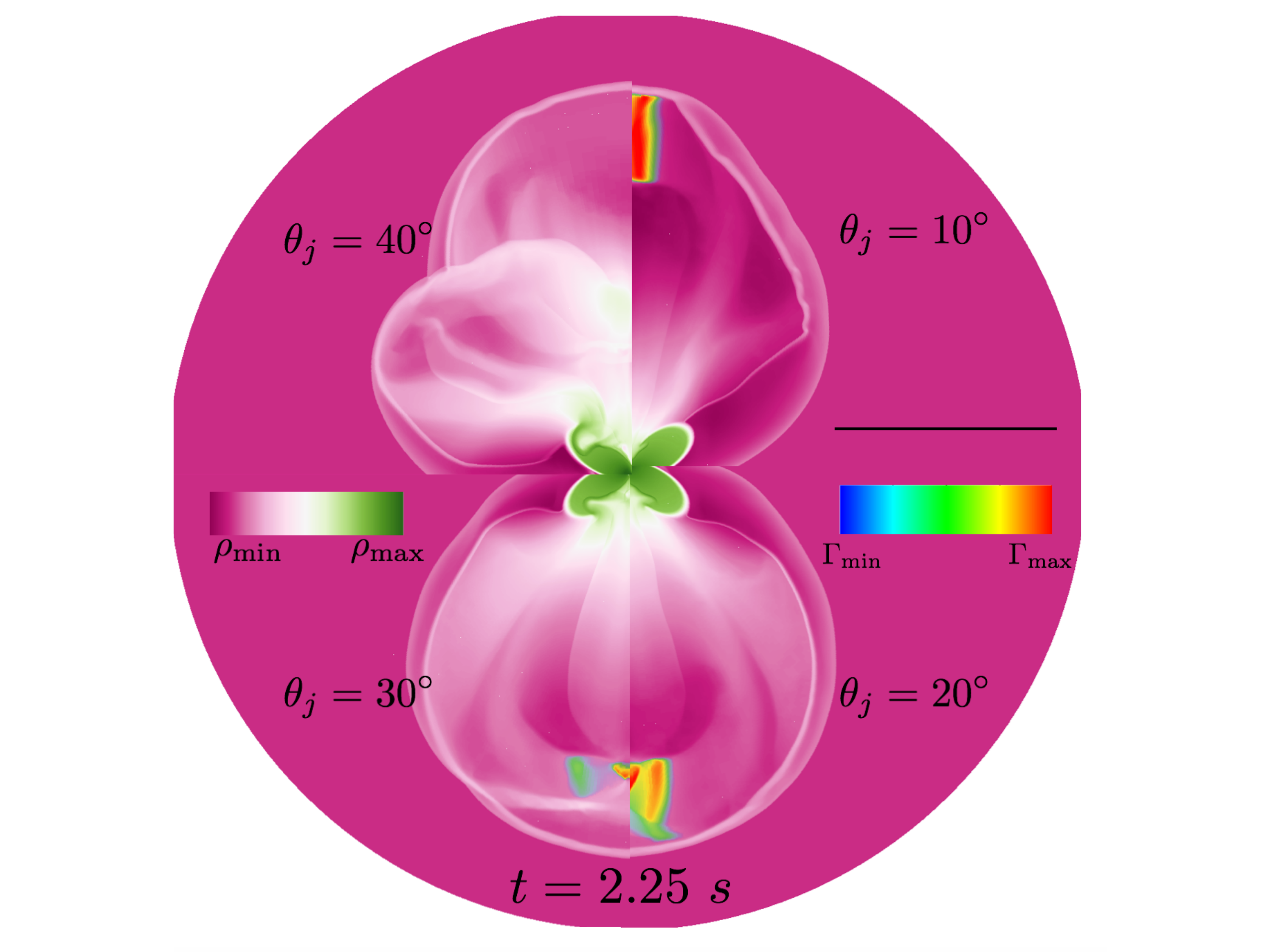}
\caption{ The structure of jets emerging from a realistic wind. The jet has $L_{\rm \Omega}=10^{51}{\rm erg\ s^{-1}}$, $\Gamma_{\rm j}=10$, $t_{\rm j}=0.5\rm{s}$ and varying $\theta_{\rm j}$. For the wind: $t_{\rm w}=0.5\rm{s}$. Shown are logarithmic density and Lorentz factor contours at time $t=2.25\rm{s}$. The range in densities: $[\rho_{\rm min}, \rho_{\rm max}]=[7.6\times10^{-7},1386.3]{\rm g\ cm^{-3}}$ and Lorentz factors: $[\Gamma_{\rm min}, \Gamma_{\rm max}]=[8.0,10.0]$. The setup is the same as in Figure~\ref{fig:numerical1}.}
\label{fig:numerical2}
\end{figure}

\begin{figure}[t!]
  \centering
\includegraphics[width=0.51\textwidth]{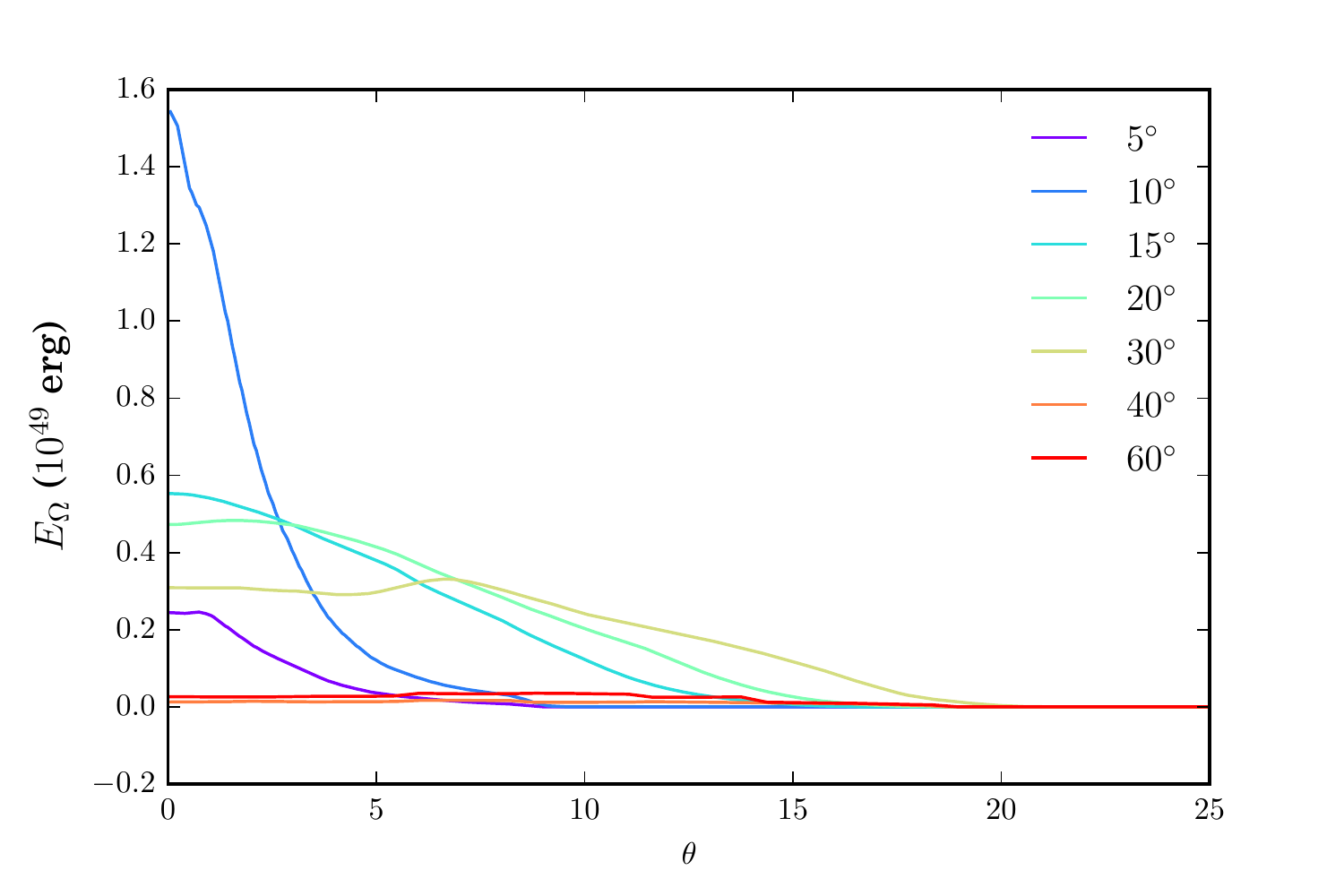}
\caption{The distribution of relativistic energy ($\Gamma \gtrsim 8$) per solid angle averaged in radius for the jets in Figure ~\ref{fig:numerical2}, for varying initial jet half-opening angle $\theta_{\rm j}$. }
\label{fig:dedomega}
\end{figure}

The angular properties of the wind also have an important effect on the jet. Figure~\ref{fig:numerical2} shows the structure of jets varying $\theta_{\rm j}$ after emerging from the wind region. For narrow jets, the outer edge of the wind is reached in a crossing time that may matter little when compared to $t_{\rm j}$. Nonetheless, wider jets have to traverse higher density regions and they may be unable to breakthrough despite having the same $L_{\rm\Omega}$.
An initially wide jet could advance much faster along the rotation axis and may eventually escape
along the direction of least resistance, getting further collimated before emerging. 
Figure~\ref{fig:dedomega} shows the relativistic energy per solid angle for the jet configurations shown in Figure~\ref{fig:numerical2}, where only narrow jets are able to successfully emerge from the wind region and, in some cases, experience significant collimation.
 
We can conclude that the properties surrounding a HMNS at the time the collapse to a BH occurs have a decisive effect on the propagation and jet's angular structure. Whether a sGRB will be observed depends sensitively not only on the power and jet's duration but also on its initial angular structure, and patchiness of the wind. Thus, the detection of a successful sGRB and its parameters provides a clear test of the neutron star merger model and the precise measurement of the duration, luminosity and jet's half-opening angle  may help constrain the dimensions and mass distribution of the HMNS region.

\section{Discussion}\label{dis}
Soon after the coalescence of two neutron stars, dense mass outflows are generated within the HMNS. The resulting wind can hamper the advancement of a relativistic jet, potentially leading to a failed sGRB. In order to break free from the wind, two conditions must be satisfied. 

First, the velocity of the jet's head has to be greater than that of the wind. For the realistic wind profiles analyzed here, we find that this simple constraint can be fulfilled if the jet power exceeds some particular limit that is found to increase with increasing opening angle.  When compared to the observed distributions of isotropic equivalent $\gamma$-ray luminosities and half-opening angles \citep{fong2015, fong2016, troja2016}, the required constraints are in agreement with observations (Figure~\ref{fig:lvstheta}), giving credence to the idea that the properties of sGRB jets are likely shaped by their interaction with the HMNS wind. 

Second, the jet's head needs to emerge from the wind's outer boundary before the central engine ceases to operate. This requirement allows us to derive a limit on the duration of the wind injection phase, which is determined by the lifetime of the HMNS. Figure~\ref{fig:timecollapse} shows the strict constraints on the time of collapse derived using the durations of sGRBs. In all instances, the required times of collapse are larger than the dynamical timescale of the HMNS but are significantly shorter than the cooling timescale of the remnant. This naturally argues against the complete stabilization of the HMNS and suggests that the collapse to a BH occurs promptly \citep{ari2014}, which in turn can be used to constrain the equation of state of nuclear matter \citep{fryer2015,cole2015}.  Because even a minuscule mass fraction of baryons polluting the sGRB jet severely limits the maximum attainable Lorentz factor, we argue that jet triggering has to be delayed until after BH collapse. 
 
In this paper, we use realistic latitudinal wind profiles arising from global simulations reproducing the conditions of NSBMs \citep{siegel2014, perego2014}, that show that the HMNS experiences non-spherical mass loss and present a detailed numerical study of the propagation of relativistic jets in such environments. Many different studies have addressed the role of external collimation \citep{rosswog03,siegel2014, perego2014,hotokezaka2013,nagakura2014}. For example, \citet{duffell2015} finds that an oblate density profile can produce collimation, relaxing the condition of an initially narrow jet. In our study we find that wider jets ($\theta_{\rm j}=40^\circ$) are likely to result in the wind choking them (Figure~\ref{fig:dedomega}), while narrow jets ($\theta_{\rm j}=10^\circ$) of similar isotropic luminosity can produce a successful sGRB. We also find that in some limiting cases, wider jets might be able to successfully break out of the wind region and emerge as more collimated outflows than initially created.

The resulting sGRB depends on the properties of the HMNS, especially the latitudinal structure of the wind. This implies that we cannot be too specific about the initial jet's structure when triggered. For example, neutrino pair annihilation is expected to produce jets with $\theta_{\rm j}\approx30^\circ$ \citep{rosswog02}, which also matches the half-opening angle of the magnetic-jet structure in NSBMs \citep{rezzolla11}, but these are likely to be further modified by the interaction with the surrounding environment. It also implies that low luminosity wider jets, can be constraining on the properties of their progenitor. The surrounding environment is expected to be less dense in the case of NS-BH progenitors. In such systems, the constraints on the properties needed for a successful event are not as stringent as in the case of NSBMs\citep{lee2004,rosswog2005, just2016}. The luminosity and opening angle of a sGRB would then provide a natural test to distinguish between different merger channels. We claim that this is the case for GRB050724 \citep{grupe2006}, in which a low luminosity, wide jet event was observed. T\citet{fongmagnetar} rules a long lived for the sGRB. Given its properties, we speculate it was likely produced by a NS-BH merger that promptly collapsed to a BH. \citet{rosswog2005} and \citet{just2016} argue that low luminosity events could be caused by neutrino pair annihilation or magnetized outflows in those types of mergers, respectively.

The task of finding useful progenitor diagnostics is simplified if the pre-burst evolution leads to a significantly enhanced wind medium in the immediate jet's vicinity. The total luminosity and opening angles observed from sGRBs are diverse. One appealing aspect of the NSBM progenitors is that the interaction with the winds of HMNS can probably explain this diversity. 
 
\section*{Acknowledgements }
We thank R. Foley, C. Fryer, M. MacLeod, J. Law-Smith, W-F. Fong, P. Kumar, C. Miller, and acknowledge financial support from the David and Lucile Packard Foundation, NSF (AST0847563), UCMEXUS (CN-12-578), and UCMEXUS-CONACYT Doctoral Fellowship. GM acknowledges  AAUW American Fellowship 2014-15. FDC acknowledges UNAM-PAPIIT grant IA103315. AP acknowledges the Helmholtz-University Investigator grant VH-NG-825. LR acknowledges ÒNewCompStarÓ, COST Action MP1304, LOEWE-Program in HIC for FAIR, European UnionÕs Horizon 2020 Research and Innovation Programme grant 671698 (FETHPC-1-2014, project ExaHyPE). SR acknowledges the Swedish Research Council (VR) grant 621-2012-4870 and  ÓNewCompStarÓ, COST Action MP1304. KT acknowledges JSPS KAKENHI grant 15H06813 and LOEWE-Program in HIC for FAIR.For analysis, we used {\it yt} analysis toolkit \citep{turkyt}.

\end{document}